%% file: project.tex
\pdfoutput=1

\documentclass[11pt]{article}
\usepackage{authblk}
\usepackage{emnlp2021}

\usepackage[T1]{fontenc}
\usepackage[utf8]{inputenc}

\usepackage{times}
\usepackage{url}
\usepackage{latexsym}
\usepackage{enumitem}
\usepackage{color, colortbl}

\usepackage[hang,flushmargin]{footmisc}

\usepackage{microtype}

\usepackage{booktabs}
\usepackage{multirow}
\usepackage{adjustbox}
\usepackage{arydshln}
\usepackage{tabularx}
\usepackage{array}
\usepackage{graphicx,subcaption}
\usepackage{enumitem}
\usepackage{amsmath}
\usepackage{amsfonts}
\usepackage{amssymb}
\usepackage{dsfont}

\definecolor{bluencs}{rgb}{0.0, 0.53, 0.74}

\newcommand\footnoteref[1]{\protected@xdef\@thefnmark{\ref{#1}}\@footnotemark}
\definecolor{Gray}{gray}{0.9}

\let\oldFootnote\footnote
\newcommand\nextToken\relax

\renewcommand\footnote[1]{%
    \oldFootnote{#1}\futurelet\nextToken\isFootnote}

\newcommand\isFootnote{%
    \ifx\footnote\nextToken\textsuperscript{,}\fi}

\title{Contextualized Query Embeddings for Conversational Search}

\author{Sheng-Chieh Lin}
\author{Jheng-Hong Yang}
\author{Jimmy Lin}
\affil{David R. Cheriton School of Computer Science\\University of Waterloo}

\begin{document}
\maketitle
\begin{abstract}
This paper describes a compact and effective model for low-latency passage retrieval in conversational search based on learned dense representations.
Prior to our work, the state-of-the-art approach uses a multi-stage pipeline comprising conversational query reformulation and information retrieval modules.
Despite its effectiveness, such a pipeline often includes multiple neural models that require long inference times.
In addition, independently optimizing each module ignores dependencies among them.
To address these shortcomings, we propose to integrate conversational query reformulation directly into a dense retrieval model.
To aid in this goal, we create a dataset with pseudo-relevance labels for conversational search to overcome the lack of training data and to explore different training strategies.
We demonstrate that our model effectively rewrites conversational queries as dense representations in conversational search and open-domain question answering datasets.
Finally, after observing that our model learns to adjust the $L_2$ norm of query token embeddings, we leverage this property for hybrid retrieval and to support error analysis. 
\end{abstract} 
\maketitle

\section{Introduction}

\begin{figure}[t]
\centering
\begin{subfigure}[t]{\columnwidth}
   \includegraphics[width=\columnwidth]{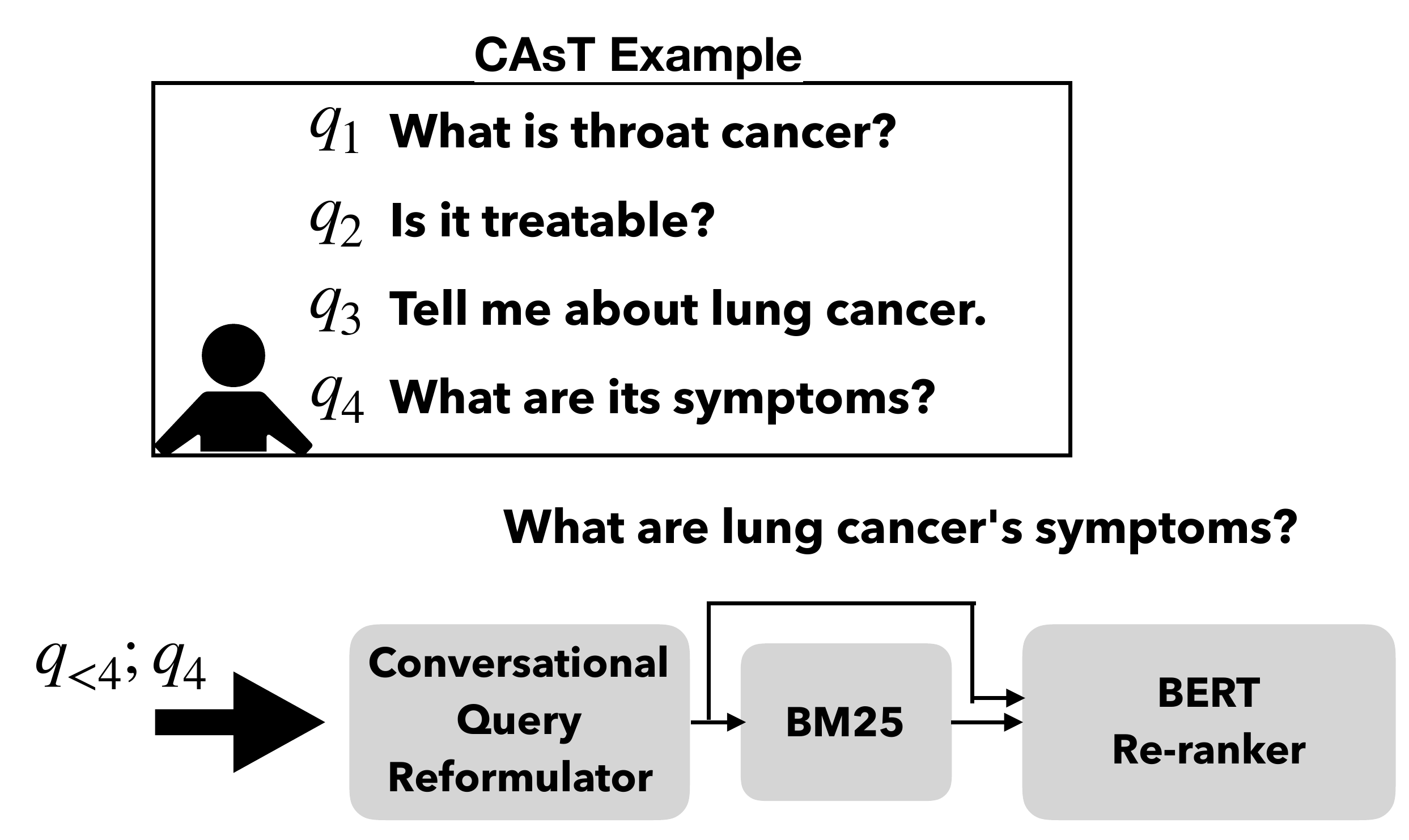}
    \caption{Multi-stage pipeline for conversational search:\ fine-tuned language models are used to reformulate user utterances for the downstream IR pipeline.}
\end{subfigure}

\vspace{0.25cm}

\begin{subfigure}[t]{\columnwidth}
    \includegraphics[width=\columnwidth]{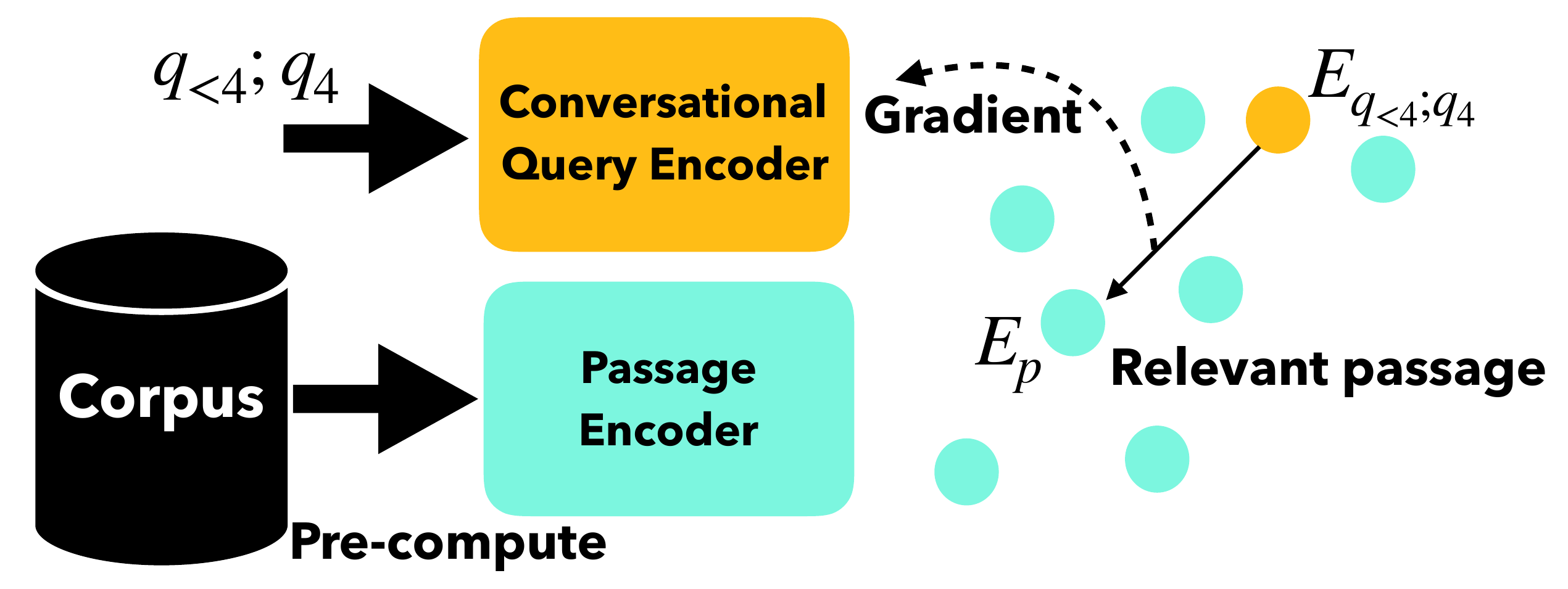}
    \caption{End-to-end conversational search:\ query reformulation is directly incorporated into the IR pipeline, thus enabling end-to-end training.}
\end{subfigure}
\caption{A comparison between (a) a multi-stage pipeline and (b) our proposed method for conversational search.}
\label{fig:illustration}
\end{figure}

With the growing popularity of virtual assistants (e.g., Alexa and Siri), information seeking through dialogues has attracted many researchers' attention. 
To facilitate research on conversational search (ConvS), \citet{cast} organized the TREC Conversational Assistance Track (CAsT) and defined ConvS as the task of iteratively retrieving passages in response to user queries in a conversation session.
An example conversation in the CAsT dataset is shown at the top of Figure~\ref{fig:illustration}(a).   

There are two main challenges for the task of conversational search:\ (1) User utterances are often ambiguous when treated as stand-alone queries since omission, coreference, and other related linguistic phenomena are common in natural human dialogues.
Hence, directly feeding the utterances into IR systems would lead to poor retrieval effectiveness.
Understanding queries through conversational context is required.
(2) There is limited data regarding conversational search for model training. 
To address the aforementioned challenges, existing papers~\cite{lin2020query, yu2020fewshot, quretec, kumar-callan-2020-making} take a multi-stage pipeline approach.
They train a conversational query reformulation (CQR) model using publicly available datasets~\cite{canard, quan-etal-2019-gecor} and feed the automatically decontextualized queries to an off-the-shelf IR pipeline~\cite{marco_BERT}.
However, such ConvS pipelines can be slow (i.e., over 10s per query on GPUs).
Furthermore, this design assumes that the reformulated queries are independent of the downstream IR pipeline, which may not be true.

In this paper, we study a low-latency end-to-end approach to ConvS.
Specifically, we adopt a bi-encoder model and incorporate CQR into the query encoder, illustrated in Figure~\ref{fig:illustration}(b).
To overcome the challenge of limited training data, we create a dataset with pseudo-relevance labels to guide the query encoder to rewrite conversational queries in latent space directly.
One may consider this approach as throwing conversational queries into a black box since the reformulated queries are represented as dense vectors.
However, we find that the fine-tuned contextualized query embeddings (CQE) are easily interpretable.
They can be transformed into text for failure analysis and can facilitate dense--sparse hybrid retrieval.

Our contributions are summarized as follows:
(1) We integrate two tasks in ConvS, query reformulation and dense passage retrieval, into our dense representation learning framework.
Due to the lack of human labeled data, we create a dataset with pseudo-relevance labels for model training.
We empirically show that our model successfully learns to reformulate conversational queries in a latent representation space.
(2) We uncover how CQE learns to reformulate conversational queries in a latent space. 
Based on this finding, we can easily transform CQE into a text (sparse) representation.
We demonstrate that the CQE text representation also performs well on sparse retrieval and can further improve CQE retrieval effectiveness using a hybrid of sparse and dense retrieval.
The CQE text also helps us understand why the technique fails or succeeds.
(3) We show that the query latency of CQE (without re-ranking) is at least an order of magnitude lower than existing multi-stage ConvS pipelines while yielding competitive retrieval effectiveness.
Hence, CQE is superior for integration with other models in downstream tasks.
(4) We empirically demonstrate its effectiveness in open-domain conversational question answering in a zero-shot setting.

\section{Preliminaries}

Let us define a sequence of conversational queries $Q = (q_{1}, \cdots, q_{i-1}, q_{i})$ for a topic-oriented session $s$, where $q_{i}$ stands for the $i$-th user query ($i \in \mathbb{N}^{+}$) in the session.
The goal of conversational search is to find the set of relevant passages ${P}^{+}_{i}$ for the user query $q_i$ at each turn, given the conversational context ${q_{<i}}$.
Thus, the task can be formulated as the objective:
\begin{equation}
\label{eq:cs}
	\mathop{\arg\max}_{\theta} \sum_{p \in P^{+}_i}\mathcal{F}^{\text{ConvS}}_{\theta}(q_{<i};q_i,p)
\end{equation}
\noindent where $\mathcal{F}^{\text{ConvS}}_{\theta}$ is the function (parameterized by $\theta$) to compute a relevance score between the conversational query $(q_{<i};q_i)$ and passage $p$.

\begin{figure*}[t!]
\centering
\begin{subfigure}[t]{2\columnwidth}
    \includegraphics[width=\columnwidth]{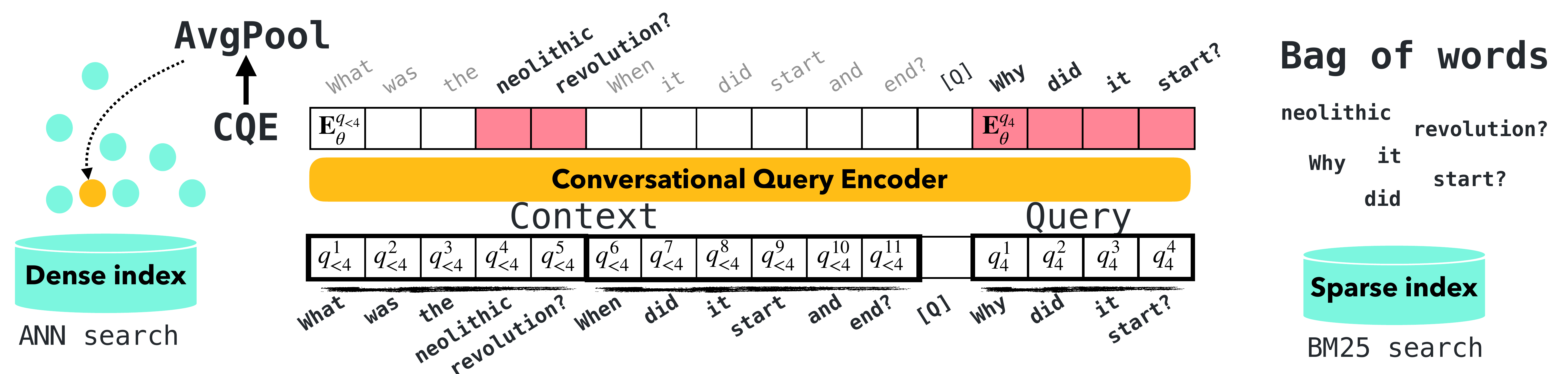}
\end{subfigure}
\caption{Our contextualized query token embeddings can be used both for dense and sparse retrieval. The left side illustrates CQE for dense retrieval by average pooling of token embeddings. The right side shows that the token embeddings can be used to select tokens from the context to form a decontextualized bag-of-words query for sparse retrieval.}
\label{fig:cqe}
\end{figure*}

Since end-to-end training data for conversational search is extremely limited, a common approach is to factorize $\mathcal{F}^{\text{ConvS}}_{\theta}$ into a multi-stage pipeline.
In a multi-stage pipeline, the components can be tuned with data collected at different stages:
\begin{equation}
\label{eq:ir}
	\mathcal{F}^{\text{ConvS}}_{\theta} \triangleq \mathcal{F}^{\text{ir}}_{\phi}(q^{*}_{i},p) \cdot \mathcal{F}^{\text{cqr}}_{\varphi}(q^{*}_{i}|q_{<i};q_i),
\end{equation}
where $q^{*}_{i}$ is the stand-alone oracle query that best represents the user's information need given the context $q_{<i};q_i$.
$\mathcal{F}^{\text{ir}}_{\phi}$ and $\mathcal{F}^{\text{cqr}}_{\varphi}$ denote the components of information retrieval (IR) and conversational query reformulation (CQR), respectively. 
Thus, Eq.~(\ref{eq:cs}) can be approximated by separately maximizing $\mathcal{F}^{\text{ir}}_{\phi}$ and $\mathcal{F}^{\text{cqr}}_{\varphi}$.
For example, we can reuse the representative \textit{ad hoc} retrieval pipeline comprised of BM25 + BERT re-ranking for $\mathcal{F}^{\text{ir}}_{\phi}$, then conduct the CQR task for $\mathcal{F}^{\text{CQR}}_{\varphi}$.

Specifically, the most common current approach~\cite{lin2020query, quretec, vakulenko2020question, kumar-callan-2020-making, yu2020fewshot} is to fine-tune a pretrained language model (LM) supervised by decontextualized queries manually rewritten by humans, and then use the fine-tuned LM to reformulate user queries for BM25 retrieval and BERT re-ranking, as illustrated in Figure~\ref{fig:illustration}(a).
While effective, this approach has two limitations:\
(1) although mimicking the way humans rewrite queries is reasonable, it hypothesizes that the optimal decontextualized queries are manually rewritten queries, which may not be true;
(2) the CQR and IR modules rely on computation-demanding pretrained LMs; thus, when combined together, they are often too slow for practical applications.

\section{Our Approach}

In this section, we first explain why a bi-encoder design is a good fit for conversational search, and then introduce our contextualized query embeddings (CQE) for conversational search.

\subsection{Bi-encoder Model}

Recently, dense passage retrieval based on bi-encoders~\cite{sentence-bert, dpr, xiong2020approximate, lin-etal-2021-batch} has attracted the attention of researchers due to its good balance between efficiency and effectiveness.
Bi-encoder models are trained to encode queries and passages in a shared latent space.
At query time, only query texts are encoded to search for the nearest passage embeddings, which are precomputed by the passage encoder.
Formally speaking, the relevance score $\phi$ of a given query $q_i$ (with its context $q_{<i}$) and a passage $p$ is computed as the dot product of their embeddings:
\begin{equation}
\label{eq:bi-encoder}
 \phi \left( (q_{<i}; q_i), p \right)
= \langle\mathbf{E}^{ (q_{<i}; q_i)}_{\theta},
\mathbf{E}^p_{\theta}\rangle,
\end{equation}
\noindent where $\mathbf{E}_{\theta}^{(\cdot)} \in \mathbb{R}^{h}$ is the BERT representation of the input texts, which can be the average or maximum pooling over token embeddings or a specific token embedding (e.g., the [CLS] embedding in BERT), and $\theta$ represents the parameters of BERT.

In this study, we adopt average pooling over token embeddings, which lets us interpret CQE easily, as we will discuss later.
Thus, we can formulate conversational search as maximizing the following log likelihood:
\begin{align}
\label{eq:nll}
&\mathcal{F}^{\text{ConvS}}_{\theta}\left((q_{<i};q_i),p\right)  \nonumber\\
&\triangleq\text{log}
\frac{\text{exp}\left(\langle\mathbf{E}^{ (q_{<i}; q_i)}_{\theta},
\mathbf{E}^p_{\theta}\rangle / \tau\right)} {\sum_{p'\in \mathcal{D}} \text{exp} \left(\langle\mathbf{E}^{ (q_{<i}; q_i)}_{\theta},
\mathbf{E}^{p'}_{\theta}\rangle / \tau\right)},
\end{align}
\noindent where $\tau$ denotes the temperature parameter and $\mathcal{D}$ is the set of passages comprising the corpus. 
In practice, $\mathcal{D}$ is replaced by the subset $\mathcal{D_B}$, consisting of the passages in a training batch, i.e., the positive and negative passages from all the queries in the same batch.
With Eq.~(\ref{eq:nll}), the optimization problem of Eq.~(\ref{eq:cs}) can be approached by end-to-end representation learning, which can be interpreted as projecting a conversational query $\mathbf{E}^{(q_{<i}; q_i)}_{\theta}$ into the latent space such that it has maximum dot product with its relevant passage $p^+$.

\subsection{Contextualized Query Embeddings}

Given the conversational context and query tokens $(q^{1}_{<i} \cdots q^{j}_{<i}, q^{1}_{i} \cdots q^{k}_{i})$, we define contextualized query embeddings (CQE) formally as $\mathbf{E}^\text{cqe}_{\theta} \in \mathbb{R}^{(j+k)\times h}$ based on BERT's contextualized token embeddings:
\begin{align}
\label{eq:cqe}
(\overbrace{\mathbf{E}^{q_<i}_{\theta}(1) \cdots \mathbf{E}^{q_<i}_{\theta}(j)}^\text{context}
, 
\overbrace{\mathbf{E}^{q_i}_{\theta}(j+1) \cdots \mathbf{E}^{q_<i}_{\theta}(j+k)}^\text{query}
).
\end{align}
Here, we take the last layer's hidden representations from BERT.
From $\mathbf{E}^{\text{cqe}}_{\theta}$, a single vector query embedding can be computed by average pooling the query token embeddings:
\begin{equation}
\label{eq:query_embedding}
\mathbf{E}^{ (q_{<i}; q_i)}_{\theta} = \frac{1}{j+k} \sum_{l=1}^{j+k} \mathbf{E}^\text{cqe}_{\theta}(l) \in \mathbb{R}^{h}.
\end{equation}
\noindent We then use $\mathbf{E}^{ (q_{<i}; q_i)}_{\theta}$ to conduct nearest neighbor search for the top-$k$ passages in the corpus using an off-the-shelf library (Facebook's Faiss, in our case), as shown in Figure~\ref{fig:cqe} (left).

\begin{figure}[t]
\centering
\begin{subfigure}[t]{1\columnwidth}
    \includegraphics[width=\columnwidth]{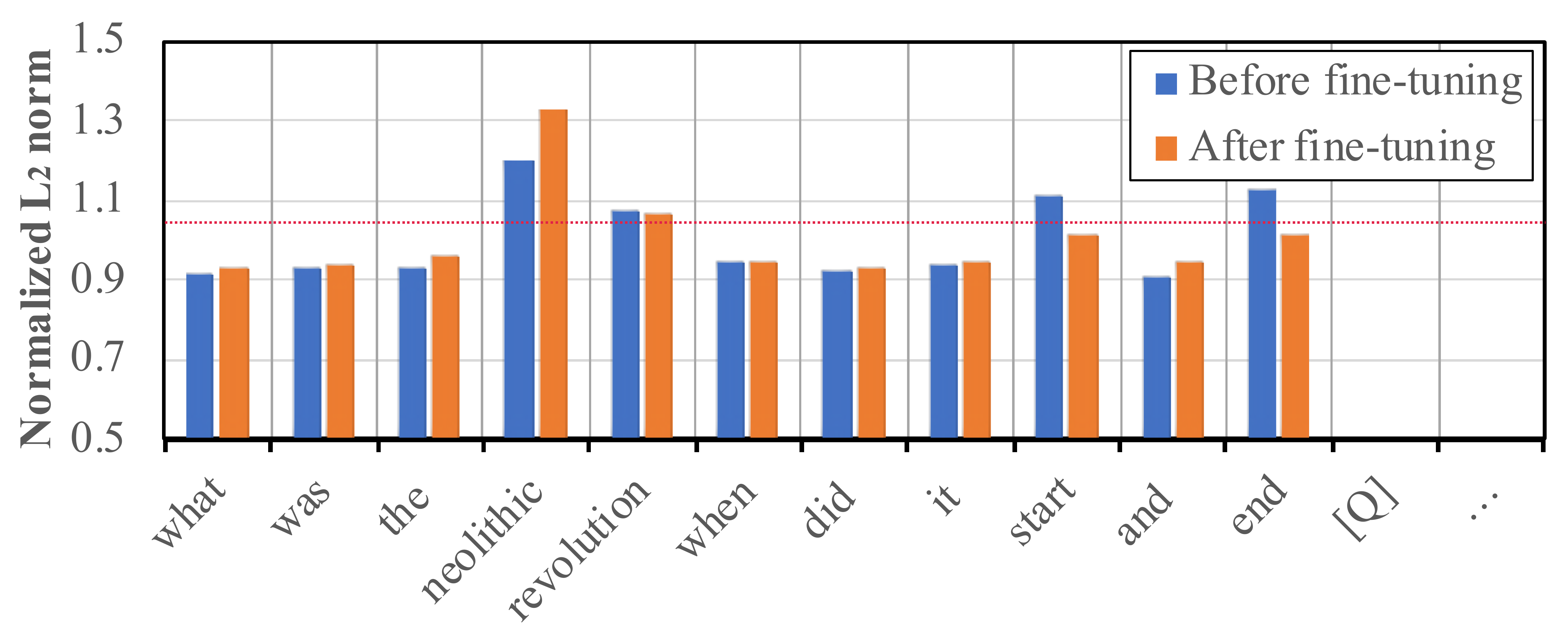}
\end{subfigure}
\caption{$L_2$ norm distribution of context token embeddings, normalized by the mean of $L_{2}$ norms among the context tokens. After fine-tuning, $L_2$ norms of context tokens that are non-relevant (e.g., start, end) decrease, and the relevant ones (e.g., neolithic) increase.}
\label{fig:interpretation}
\end{figure}

\subsection{Interpreting CQE}
\label{subsec:interpret}

While condensing a multi-stage pipeline into single-stage dense retrieval is attractive, it may be difficult for interpretation (i.e., we cannot examine the reformulated queries). 
In this subsection, we explain how to interpret CQE.
With Eq.~(\ref{eq:query_embedding}), we rewrite Eq.~(\ref{eq:bi-encoder}) as the average dot product of each token embedding $\mathbf{E}^\text{cqe}_{\theta}(l)$ and a single-vector passage embedding $\mathbf{E}^p_{\theta}$:
\begin{align}\label{eq:interpret}
&\phi \left( (q_{<i}; q_i), p \right)
\nonumber\\
&=\frac{1}{j+k} \sum_{l=1}^{j+k} \|\mathbf{E}^\text{cqe}_{\theta}(l)\|_2 \langle \hat{\mathbf{E}}^\text{cqe}_{\theta}(l) , \mathbf{E}^p_{\theta} \rangle,
\end{align}
\noindent where $\hat{\mathbf{E}}^\text{cqe}_{\theta}(l)$ is a unit vector.
Intuitively, to maximize Eq.~(\ref{eq:nll}), CQE can learn to adjust the $L_2$ norm of $\mathbf{E}^\text{cqe}_{\theta}(l)$ when we freeze the passage embeddings.
To be more specific, it appears that CQE learns to increase the $L_2$ norm for relevant query--passage pairs and decrease it otherwise.
Thus, we can consider the $L_2$ norm of each token embedding as its term importance for retrieving relevant passages.

For the example in Figure~\ref{fig:cqe}, we empirically analyze the query token embeddings of our CQE model.
Figure~\ref{fig:interpretation} shows the normalized $L_2$ norm for the context of the user query (``\textit{why did it start?}'').
We observe that after fine-tuning, the terms ``\textit{neolithic}'' and ``\textit{revolution}'' show greater $L_2$ norms than the others.
On the other hand, the $L_2$ norms for the terms ``\textit{start}'' and ``\textit{end}'' decrease.

With this observation, we can use CQE to generate decontextualized queries. 
Specifically, inspired by the term weighting ideas of~\citet{deepct}, we conduct query expansion by selecting the terms ($\|\mathbf{E}^{q_<i}_{\theta}(\cdot)\|_2 \geq \gamma$, where $\gamma$ is a hyperparameter) from the context using CQE and concatenate them to the user query $q_i$, illustrated on the right side of Figure~\ref{fig:cqe}.
Note that the decontextualized queries generated by CQE are bag-of-words sets rather than fluent natural language queries.
However, in Section~\ref{sec:exp}, we show that the decontextualized queries can be used for sparse retrieval and even for conducting failure analysis.

\section{Training Data and Strategies}

In this section, we first introduce how we create weakly supervised training data for conversational search.
Then, we discuss some possible strategies to fine-tune CQE.

\paragraph{Weakly supervised training data.}
By taking the idea of pseudo-labeling, we create our weakly supervised training data for end-to-end conversational search.
There are human rewritten queries that help models learn to decontextualize them in conversation;
however, only limited labels are available for end-to-end conversational search, as shown in Table~\ref{tb:cast_stat}.
Hence, we combine three existing resources to train our model with weak supervision:\ (1) CANARD~\cite{canard}, a conversational query reformulation dataset; (2) ColBERT~\cite{colbert}, a strong text ranking model trained on MS MARCO for passage ranking; and (3) the passage collection provided by the TREC CAsT Tracks~\cite{cast}.

To combine the three resources, we make a simple assumption: decontextualized queries can be paired with their relevant passages selected by ``good enough'' \textit{ad hoc} retrieval models.
Thus, for each human reformulated query in the CANARD dataset, we retrieve 1000 candidate passages from the CAsT collection using BM25, and then re-rank them using ColBERT.
We assume the top-3 passages are relevant for each query, while treating the rest as non-relevant.

\begin{table}[t]
\centering
      \caption{CANARD dataset statistics.}
      \label{tb:canard_stat}
      \begin{small}
        \begin{tabular}{lrrr}
          \toprule
	     CANARD & Training & Dev & Test \\
        \midrule
     \# Queries & 31,526  & 3,430 &5,571 \\
    \# Dialogues & 4,383  & 490 & 771  \\
     \bottomrule
        \end{tabular}
        \end{small}
\end{table}

\begin{table}[t]
\caption{CAsT dataset statistics.}
\centering
\label{tb:cast_stat}
\resizebox{.75\columnwidth}{!}{
    \begin{tabular}{lrrr}
       \toprule
       & \multicolumn{2}{c}{CAsT19} & CAsT20 \\
       \cmidrule(lr){2-3} \cmidrule(lr){4-4}
       & Training & Eval & Eval\\
    \midrule
 \# Queries & 108  & 173 & 208 \\
\# Dialogues & 13  & 20 & 25   \\
\cdashline{1-4}
\# Passages & \multicolumn{3}{c}{38M} \\
 \bottomrule
    \end{tabular}
    }
\end{table}

\paragraph{Bi-encoder warm-up.}
Training a bi-encoder model for dense retrieval requires lots of data, not to speak of conversational search.
Following previous work on conversational search~\cite{yu2020fewshot,lin2020query,vakulenko2020question}, we adopt MS MARCO as our bi-encoder warm-up training dataset, where the training procedure is adopted from the work of~\citet{lin2020distilling}.

\paragraph{CQE fine-tuning.}
After bi-encoder warm-up, we fine-tune the \textit{query} encoder to consume conversational queries and generate contextualized query embeddings. 
Specifically, for each query $q_i$ in our training data, we sample a triplet $([q_{<i};q_i], p^{+}, p^{-})$ for fine-tuning, where $p^{+}$ and $p^{-}$ are  sampled from positive passages (labeled by ColBERT) and top-200 BM25 passages (without replacement), respectively.
Note that, at this stage, we freeze our passage encoder and only fine-tune the query encoder; thus, we can precompute all the passage embeddings in the CAsT corpus, and only encode queries for evaluation.
In this work, we further explore different strategies to better train CQE using our weakly supervised training data.

\paragraph{Hard negative mining.}
Although sampling negatives from BM25 top-$k$ candidates is effective for dense retrieval training, \citet{xiong2020approximate} demonstrate that hard negatives bring more useful information for training dense retrievers.
In this work, we explore whether hard negatives benefit the fine-tuning of our CQE model.
Instead of using asynchronous index refreshing, as in the work of \citet{xiong2020approximate}, we sample hard negatives $p^-$ from the top-$200$ passages re-ranked by ColBERT.

\paragraph{Training with soft labels.}
Due to the strong assumptions we make for weak supervision, using cross entropy for one-hot pseudo-label training may be sub-optimal because our model could be overconfident about its predictions.
To address this issue, we use the logits of ColBERT as soft labels to fine-tune CQE to have similar confidence predictions, i.e., knowledge distillation.
It is worth noting that we only minimize the KL divergence of softmax normalized dot products with respect to \text{in-batch} query--passage pairs without using cross entropy for interpolation, as in the traditional (strongly) supervised setting.

\section{Experimental Setup}\label{sec:exp}

\paragraph{Datasets.} 
We conduct experiments on TREC CAsT datasets.
TREC organized the Conversational Assistance Tracks~\cite{cast}, aiming to collect reusable collections for conversational search.
The organizers have created relevance judgments (relevance grades 0--4) for each query using assessment pools from participants.
In total, there are three datasets available, CAsT19 (training and eval) and CAsT20 (eval).\footnote{\url{https://github.com/daltonj/treccastweb}} 
The dataset statistics are listed in Table~\ref{tb:cast_stat}.
All relevant passages come from the CAsT corpus (consisting of 38M passages). 
In addition, we demonstrate the generalization of CQE on an open-domain conversational question answering dataset ORConvQA in a zero-shot setting, detailed in Section~\ref{sec:orconvqa}.

\begin{table*}[t]
	\caption{CAsT passage retrieval effectiveness comparisons. 
	The best automatic approach in the same comparison group (sparse/dense) is bolded. Superscripts denote significant differences with respect to CQE (paired $t$-test $p<0.05$); $^{*}$ denotes significant difference between CQE-hybrid and all the other automatic approaches.
	W/T denotes \# of queries win/tie against human queries.}
	\label{tb:dense_retrieval}
	\centering
	\resizebox{2.05\columnwidth}{!}{
	\setlength{\tabcolsep}{4pt}
    \begin{tabular}{llccllccllcc}
	\toprule
	\multicolumn{2}{l}{} &\multicolumn{2}{c}{\# Params  \ \ Latency} &\multicolumn{4}{c}{CAsT19 Eval}& \multicolumn{4}{c}{CAsT20 Eval} \\
 \cmidrule(lr){3-3}\cmidrule(lr){4-4}\cmidrule(lr){5-8}\cmidrule(lr){9-12}
 &Query & millions &  ms/q &  Recall  & nDCG  & nDCG@3 &W / T & Recall  & nDCG  & nDCG@3 & W / T\\
	   \midrule
          \multirow{5}{*}{\rotatebox[origin=c]{90}{Sparse}} & {(0) Humans}  & -  & -& .803 &  .510 &.309 & - &.707   & .423  & .240 &- \\
         &
	    (1) Few-Shot Rewriter  & 355& 582 & .717 & .438 & .248 & 10 /126 & .490 & .284 & .145 & 23 /121\\
	   &
	    (2) QuReTeC &  340& \ \ 29 & .768  &   \textbf{.485}  & \textbf{.296} & 27 /117 &.508  &.291 &.136 & 18 /115\\
          & {(3) NTR (T5)} & 220 & 307  & .753&    .471  & .295 & 15 /136 & .514&  .303 & .159 & 24 /115 \\
          & (4) CQE-sparse  & 110 & \ \ 11  & \textbf{.773}$^{1\ \ \ \ }$  &   .462   & .272 &\textbf{45} /\ \ 67 & \textbf{.582}$^{1-3}$    & \textbf{.332}$^{1-2}$  & \ \ \textbf{.172}$^{2}$  & \textbf{38} /\ \ 89\\
    \midrule
          \multirow{5}{*}{\rotatebox[origin=c]{90}{Dense}} & {(0) Humans}& - &- & .797 & .571 & .507 & - & .804 & .558 &.460 & -\\
         &(1) Few-Shot Rewriter &  465 & 593 & .723& .510 & .449 & 18 /117 & .611 & .378 & .256 & 10 /108 \\
          &(2) QuReTeC & 450& \ \ 40 & .773 & .545 & .473 & 38 /\ \ 82 &.600   &.390 & .288 & 26 /\ \ 78\\
         &(3) NTR (T5) & 330 & 318 & .762   & .543 & .495 & 23 /125 & .635 & .421  & \textbf{.323} & 28 /\ \ 90\\
         &(4) CQE & 110& \ \ 11 & \textbf{.784}$^{1}$  & \textbf{.559}$^{1\ \ \ \ }$  & \textbf{.499}$^{1}$ & \textbf{60} /\ \ 45 & \textbf{.699}$^{1-3}$  & \textbf{.447}$^{1-3}$ & \ \ \ \ \ \  .312$^{1-2}$ & \textbf{40} /\ \ 46 \\
    \midrule
    \midrule
      &CQE-hybrid  & 110& \ \ 11 & .823$^{*}$  & .598$^{*}$ & .515 & - & .730$^{*}$  & .475$^{*}$ & .338 &-\\
	\bottomrule
	\end{tabular}}
\end{table*}

\paragraph{Query reformulation baselines.}

A reasonable setting to compare CQE with existing query reformulation models is to directly feed the reformulated queries into a bi-encoder model for dense retrieval. 
For a fair comparison, we encode the reformulated queries into query embeddings using our pretrained bi-encoder model, which is suitable for stand-alone queries.
Note that the passage embeddings in the corpus for CQE and the other models are the same since we freeze the passage encoder while fine-tuning CQE. 
We compare CQE with three state-of-the-art conversational query reformulation models and a human baseline, described below:

\begin{itemize}[leftmargin=*]
\item \textbf{Few-Shot Rewriter}: \citet{yu2020fewshot} fine-tune the pretrained sequence-to-sequence LM GPT2-medium on CAsT manually reformulated queries and synthetic queries created by a rule base. For the CAsT19 and CAsT20 eval sets, we directly use their publicly released queries.\footnote{\label{query_source} \href{https://github.com/thunlp/ConversationQueryRewriter}{Few-Shot Rewriter},
\href{https://github.com/nickvosk/sigir2020-query-resolution}{QuReTeC}, \href{https://huggingface.co/castorini/t5-base-canard}{NTR (T5)}}

\item \textbf{QuReTeC}: \citet{quretec} conduct query expansion using BERT-large as a term classifier, which is fine-tuned on the CANARD dataset. We directly use the reformulated queries provided by the authors.\footref{query_source}

\item \textbf{NTR (T5)}: \citet{lin2020query} fine-tune the pretrained sequence-to-sequence LM, T5-base, on the CANARD dataset.
Following their work, we use the released model\footref{query_source} with beam-search inference (setting width to 10).

\item \textbf{Humans}: We also conduct experiments on the manually reformulated queries provided by the TREC CAsT organizers as a reference.

\end{itemize}

\noindent Since CQE can be used to decontextualize conversational queries, as discussed in Section~\ref{subsec:interpret}, we also apply CQE reformulated queries (denoted CQE-sparse) to sparse retrieval (after conversion to text).
The optimal $L_2$ threshold $\gamma$ (10.5) is tuned on the CAsT19 training set.

\paragraph{Model details.}
We fine-tune CQE using BERT-base for 10K steps with batch size 96 and learning rate $7\times 10^{-6}$ on all queries (training, dev, and test) in the CANARD dataset (see Table~\ref{tb:canard_stat}), and use the CAsT19 training set as our development set.
In our main experiments, we use our best training strategy, combining hard negative mining and soft labeling (see the ablation study in Section~\ref{sec:ablation}).
We perform dense retrieval using Faiss~\cite{JDH17} (Faiss-GPU, brute force) and sparse retrieval using Pyserini~\cite{Lin_etal_SIGIR2021_Pyserini} (BM25, $k_1=0.82,b=0.68$).
In addition, we measure the latency of conversational query reformulation for each model (Latency).
For CQE, we report the latency of generating the contextualized query embeddings.
Note that for encoder-only models (BERT), we set maximum input length to 150, while for decoder-only and encoder-decoder models (GPT and T5), we further set maximum output length to 32 and use greedy search decoding.
All latency measurements are from Google Colab using a single GPU (12GB NVIDIA Tesla K80).
Finally, we report the size of each model (\# Params).

\paragraph{Evaluation metrics.}
Following \citet{cast2020}, for each approach, we compare overall retrieval effectiveness using nDCG and recall (at cutoff 1000), and top-ranking accuracy using nDCG@3.
For recall, we take relevance grade $\geq 2$ as positive.
The evaluation is conducted using the \texttt{trec\_eval} tool.
In addition, for each model, we report the number of queries win (tie) against manual queries on nDCG@3.
All significance tests are conducted with paired $t$-tests ($p<0.05$).

\begin{figure*}[t]
    \includegraphics[width=1.\textwidth]{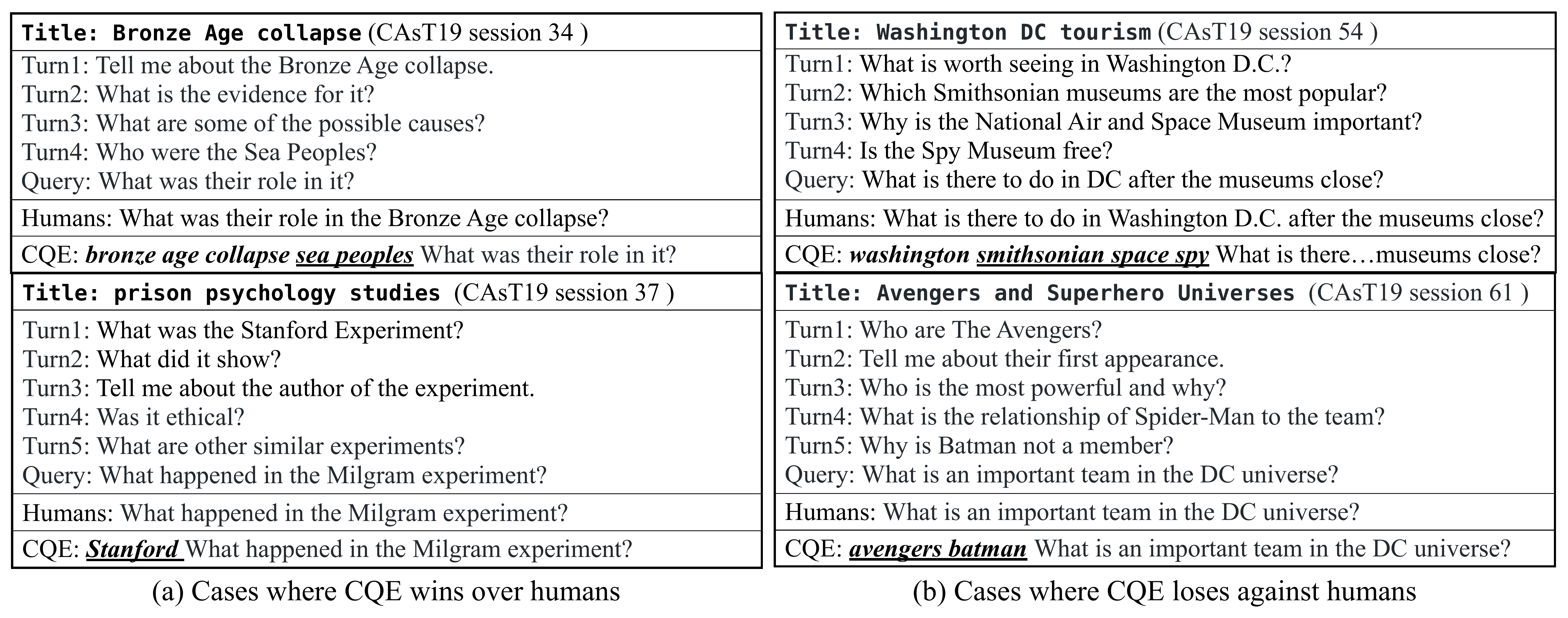}
\caption{Case studies. We choose cases based on nDCG dense retrieval scores; the CQE text shown is for sparse retrieval. Underline denotes terms {\it not} appearing in human queries.}
\label{fig:case_study}
\end{figure*}

\section{Results}

\subsection{Results on CAsT}\label{sec:result}

\paragraph{First-stage retrieval comparisons.}
Table~\ref{tb:dense_retrieval} reports the sparse and dense retrieval effectiveness of various methods. 
Overall, dense retrieval yields better effectiveness than BM25 retrieval.
Observing the first block in Table~\ref{tb:dense_retrieval}, CQE-sparse yields reasonable effectiveness compared to the other CQR models, indicating that CQE can be well represented with text.
As for dense retrieval, CQE is able to beat the other CQR models.
Although NTR (T5) and CQE yield comparable top-ranking accuracy, it is worth mentioning that unlike CQE, the other CQR modules are built independently.
Thus, when incorporated with dense retrieval, the overall memory and latency required increase, i.e., \# params of NTR (T5) increases from 220M to 330M and is much slower.

Finally, we also conduct CQE dense--sparse hybrid retrieval using their linear score combination (denoted by CQE-hybrid); see Appendix A for detailed settings.
CQE-hybrid retrieval effectiveness shows significant gains over CQE dense only.
The gains from the dense--sparse hybrid suggest that the textual interpretation of CQE not only helps us understand the query reformulation mechanism in dense retrieval but also improves effectiveness, all using a single, unified model.

A comparison of win (tie) entries shows that CQE has more wins against human queries than all the other CQR models.
On the other hand, the other CQR models have relatively more ties against human queries than CQE.
The difference between the queries is probably because CQE learns to reformulate conversational queries through the guide of pseudo-relevant passages, meaning that CQE approaches the task in a different way from the other CQR models, which are trained to mimic the way humans reformulate queries. 
This observation indicates that CQE provides different ``views'' from other CQR models and could further benefit from fusion with state-of-the-art CQR models, which we demonstrate in Appendix B. 

\begin{table}[t]
\caption{Comparisons to SOTA multi-stage pipelines.}
\label{tb:top-ranking}
\centering
    \resizebox{1.\columnwidth}{!}{
    \begin{tabular}{lcc}
    \toprule
	     CAsT19 Eval & nDCG@3  \\
        \midrule
    \multicolumn{2}{l}{\textbf {BERT-base}: latency = 314 ms}  \\
    CQE  &.499 \\
    CQE-hybrid  &.515\\[0.75ex]
\cdashline{1-2}\noalign{\vskip 0.75ex}
    \multicolumn{2}{l}{\textbf {CQR + BM25 + BERT-base:} latency = 5,350 ms} \\
    QuReTec~\cite{quretec}   &.476 \\
    Few-Shot Rewriter~\cite{yu2020fewshot}  &.492 \\
    \multicolumn{2}{l}{\textbf {3CQR + BM25 + BERT-base:} latency = 8,025 ms (est.)}\\
    MVR~\cite{kumar-callan-2020-making}&  .565 \\[0.75ex]
    \cdashline{1-2}\noalign{\vskip 0.75ex}  
    \multicolumn{1}{l}{\textbf {CQR + BM25 + BERT-large:} latency = 16,450 ms}\\
    Transformer++~\cite{vakulenko2020question} & .529 \\  
    NTR (T5)~\cite{lin2020query}  &.556 \\
    {HQE + NTR (T5)}~\cite{lin2020query}  &{.565} \\
     \bottomrule
	\end{tabular}
	}
\end{table}

\paragraph{Multi-stage pipeline comparisons.}
We compare our CQE method with other multi-stage pipelines in terms of top-ranking effectiveness, reported in Table~\ref{tb:top-ranking}.
All of these pipelines consist of conversational query reformulators (CQR), BM25 retrieval, and BERT re-ranking.
Here, we also list systems that use a BERT-large re-ranker for reference.
As for the retrieval latency, since the CAsT corpus requires 55 GiB for the dense vector index, we measure the latency of CQE on two V100 GPUs.
For the other BERT re-ranking pipelines, we divide the numbers reported in \citet{colbert}, which is measured on a single V100 GPU, by two for a fair comparison. 
We observe that {\it single-stage} CQE (with much lower latency) can compete with all the multi-stage pipelines that use a BERT-base re-ranker, except for MVR, which fuses three re-ranked lists from three different neural CQR models.
As expected, re-ranking with BERT-large can yield higher effectiveness, but is also much slower.
Of course, we can take CQE results and further re-rank them also.

\paragraph{Case studies.}
We demonstrate how CQE reformulates queries by comparing CQE and human reformulated queries on the CAsT19 eval set.
Figure~\ref{fig:case_study}(a) shows cases where CQE beats humans in terms of nDCG (in the dense retrieval setting).
The first example shows that humans mistakenly rewrite the query by omitting ``\emph{sea peoples}'' in the context. 
The second example shows that humans reformulate the query correctly; however, CQE further adds the key term ``\emph{Stanford}'' to the original query and obtains better ranking accuracy.
These cases tell us that manually reformulated queries may not be optimal for the downstream IR pipeline, and CQE can actually do better.
On the other hand, Figure~\ref{fig:case_study}(b) illustrates cases where CQE performs worse than humans.
In both cases, we observe that CQE adds related terms (i.e., ``\textit{avengers}'' and ``\textit{batman}''), but these terms degrade retrieval effectiveness.
This suggests that a better negative sampling strategy may be required to guide CQE to select key terms and generate more accurate embeddings under such challenging contexts.

\subsection{Zero-shot Transfer to ORConvQA}
\label{sec:orconvqa}

In this section, we examine the effectiveness of CQE on a downstream task:\ open-domain conversational question answering. 
The ORConvQA dataset is built on QuAC~\cite{quac}, a conversational question answering (ConvQA) dataset.
To better approximate open-domain ConvQA, \citet{orconvqa} share an extensive corpus using the English Wikipedia dump from Oct.\ 20th,  2019.
Then, they split 5.9 million Wikipedia articles into passage chunks with at most 384 BERT WordPiece tokens, resulting in a corpus of 11M passages.
Thus, the task is to first retrieve passages from the corpus using conversational queries and then extract answer spans from the retrieved passages.
Since the task shares the same conversational queries as our created dataset (both are built on CANARD), we fine-tune CQE only on the training set listed in Table~\ref{tb:canard_stat}. 
For a fair comparison between the retrievers, we directly use the reader provided by~\citet{orconvqa},\footnote{\url{https://github.com/prdwb/orconvqa-release}} which extracts the answer span from the top-5 retrieved passages. 

\begin{table}[t]
\caption{Results on the ORConvQA eval set.}
\label{tb:orqa}
\centering
    \resizebox{1.\columnwidth}{!}{
    \begin{tabular}{lccc}
    \toprule
       &\multicolumn{2}{c}{Retriever} & \multicolumn{1}{c}{Reader}\\
     \cmidrule(lr){2-3} \cmidrule(lr){4-4}
	     Model  & Recall & MRR & F$_1$ \\
        \midrule
    BERTserini~\cite{bertserini} & .251 & .178 &  26.0 \\ 
    \citet{orconvqa}  & .314 & .225 & 29.4  \\[0.75ex]
    \cdashline{1-4}\noalign{\vskip 0.75ex}
    CQE  & .365 & .266  & 30.5 \\
     CQE-hybrid   & .415 &.310 & 32.0\\
     \bottomrule
	\end{tabular}
	}
\end{table}

We first compare our CQE retrieval effectiveness to baselines, where the numbers are from \citet{orconvqa}.
To fairly compare with the dense retriever of \citet{orconvqa} (with 128 dimensions), we first conduct unsupervised dimensionality reduction using Faiss (OPQ128, 1VF1, PQ128) from 768 to 128 dimensions.
As shown in Table~\ref{tb:orqa}, CQE beats the other models in terms of retrieval effectiveness.
It is worth noting that the baselines are fine-tuned on ORConvQA, with the passages containing answer spans as positives.
In contrast, CQE is only fine-tuned on our weakly supervised training data.
This difference suggests that CQE has a degree of generalization capability.
More importantly, we observe that the retrieval effectiveness gain from CQE directly benefits F$_1$ scores.
Finally, with hybrid retrieval, a unique feature of CQE, we further improve both retrieval and the downstream task. 

\subsection{Fine-Tuning Ablation}
\label{sec:ablation}

We explore different training strategies with our weakly supervised training data.
We use the CAsT19 training set for evaluation and the results are reported in Table~\ref{tb:ablation}. 
Recall that, while training without hard negative sampling, we use the negatives randomly sampled from BM25 top-200 candidates.
First, we observe that simply training with our pseudo-labeled data can effectively guide model training; see condition (1) vs.~(0).
In addition, training with ColBERT's soft labels brings substantial effectiveness gains, as shown by condition (4) vs.~(3) and condition (2) vs.~(1).
Finally, although hard negative samples cannot directly enhance CQE's retrieval effectiveness, from condition (3) vs.~(1), by combining soft labels, a modest effectiveness gain can still be observed, from condition (4) vs.~(2). 
Thus, the best strategy for fine-tuning CQE on our weakly supervised training data is to combine hard negative sampling and soft labeling.

\begin{table}[t]
	\caption{Ablation study on CAsT19.}
	\label{tb:ablation}
	\centering
	\small
\resizebox{.9\columnwidth}{!}{
    \begin{tabular}{cccll}
	\toprule
  &\multicolumn{2}{c}{Strategy} & \multicolumn{2}{c}{CAsT19 Train}\\
  \cmidrule(lr){2-3} \cmidrule(lr){4-5}
	Cond. & Soft label & Hard neg. & Recall  & nDCG \\
	\midrule
	(0)& \multicolumn{2}{c}{No training} &.160  & .050\\[0.75ex]
	\cdashline{1-5}\noalign{\vskip 0.75ex}
    (1)& & & .670  & .300\\
    (2)&$\checkmark$ &  & .723  & .344\\[0.75ex]
    \cdashline{1-5}\noalign{\vskip 0.75ex}
    (3)& &$\checkmark$ &.660   &.306\\ 
    (4)& $\checkmark$ & $\checkmark$ & \textbf{.734}   & \textbf{.353} \\
	\bottomrule
	\end{tabular}
	}
\end{table}

\section{Related Work}

\paragraph{Conversational search.}
\citet{conv_search} define conversational search as addressing users' information needs through multi-turn conversational interactions, which can be classified into two scenarios:\
(1) A user is searching for a single item through multi-turn query clarifications, which has been studied by \citet{Aliannejadi2019AskingCQ, uddin2018multitask, guided_transformer}.
(2) A user is searching for multiple items surrounding a topic.
For example, when planning a vacation, a user would query some source of knowledge (possibly, even a human expert) to find information about destinations, hotels, transportation, etc.\ through conversational interactions.
Our work belongs to the latter search scenario.

\paragraph{Query reformulation.}
TREC organizers have built standard benchmark datasets, CAsT~\cite{cast}, to facilitate research on conversational search. 
Existing work built on CAsT mainly focuses on conversational query reformulation, previously studied by \citet{cqu,scaling}. 
For example, \citet{ilips,hqae} perform rule-based query expansion from dialogue context. 
\citet{yu2020fewshot, quretec, vakulenko2020question, lin2020query} fine-tune pretrained language models to mimic the way humans rewrite conversational queries.
These papers demonstrate that building a CQR model on top of IR systems works well.
However, \citet{lin2020query, kumar-callan-2020-making} point out that human reformulated queries may not be optimal for downstream IR modules.
They further address this problem by fusing the ranked lists retrieved using different CQR models; however, these solutions still rely on multi-stage pipelines.
In contrast, this work explores a \textit{single-stage}, \textit{end-to-end} approach to conversational search.

\paragraph{Conversational question answering.}
Another related thread of research is conversational question answering (ConvQA)~\cite{coqa,quac}, a downstream task of conversational search.
Most related work~\cite{hae, ahs} focuses on improving answer span extraction using dialogue context information.
\citet{orconvqa} first create an open-domain ConvQA dataset on top of QuAC~\cite{quac} and then tackle this dataset with a pipeline consisting of a retriever and a reader.
In this work, we demonstrate that weakly supervised CQE can directly serve as a strong retriever without further fine-tuning, and it improves the accuracy of answer span extraction.
Furthermore, different from~\citet{orconvqa}, CQE provides a single model that supports dense--sparse hybrid retrieval for conversational search, which further improves retrieval effectiveness. 

\section{Conclusions}

In this paper, we study how to simplify the multi-stage pipeline for conversational search and propose to integrate modules for conversational query reformulation (CQR) and dense passage retrieval into our dense representation learning framework.
To address the lack of training data for conversational search, we create a dataset with pseudo-relevance labels and explore different training strategies on this dataset.
Experiments demonstrate that our model learns to reformulate conversational queries in a latent space and generates contextualized query embeddings (CQE) for conversational search.
In addition, our analyses provide insight into how CQE learns to rewrite conversational queries in this latent space. 
Finally, we show that there are two main advantages of CQE:
First, the effectiveness of CQE is on par with state-of-the-art multi-stage pipelines for conversational search, but with much lower query latency.
Second, CQE serves as a strong dense retriever for open-domain conversational question answering.

\paragraph{Limitations and future work.}
Our work shows the feasibility of integrating conversational query reformulation and \textit{ad hoc} retrieval into a bi-encoder dense representation learning framework. 
However, it is unclear whether the same strategy can be applied to a cross-encoder re-ranker, which, although much slower, still achieves the highest levels of effectiveness. 
Another limitation of our work is that only historical queries are considered as context; nevertheless, in a real conversational scenario, other types of contexts should also be considered, e.g., system responses and conversations between multiple speakers (if present).
There is still much to explore around dense representations in these scenarios, which we leave to future work.
Finally, as shown in \citet{clear}, incorporating sparse retrieval signals into the training of dense retrieval improves dense--sparse fusion effectiveness.
We suspect that there is more to be gained from better fusion of dense and sparse results for conversational search.

\section*{Acknowledgements}

This research was supported in part by the Canada First Research Excellence Fund and the Natural Sciences and Engineering Research Council (NSERC) of Canada.
Additionally, we would like to thank the support of Cloud TPUs from Google's TPU Research Cloud (TRC).

\bibliographystyle{acl_natbib}
\bibliography{project.bib}

\clearpage

\appendix
\input{appendix}

\end{document}

%% file: appendix.tex
\section{Dense--Sparse Hybrid Settings}

For each query $q$, we use sparse and dense representations to retrieve top-1000 passages, $\mathcal{D}_{sp}$ and $\mathcal{D}_{ds}$, with their relevance scores, $\phi_{sp}(q, p \in \mathcal{D}_{sp})$ and $\phi_{ds}(q, p \in \mathcal{D}_{ds})$, respectively.
Then, we compute the score for each retrieved passage, $p \in \mathcal{D}_{sp}\cup \mathcal{D}_{ds}$, as follows:
\begin{equation}
\label{eq:combination}
\small
\phi(q,p)=
\begin{cases} 
    \alpha \cdot \phi_{sp}(q,p) +  \underset{p \in \mathcal{D}_{ds}}\min_{}\phi_{ds}(q,p), &\text{if }p \notin D_{ds} \\
     \alpha \cdot \underset{p \in \mathcal{D}_{sp}}\min_{}\phi_{sp}(q,p) + \phi_{ds}(q,p), & \text{if }p \notin D_{sp} \\
      \alpha \cdot \phi_{sp}(q,p) + \phi_{ds}(q,p), & \text{otherwise.}\\
\end{cases}
\end{equation}
Eq.~(\ref{eq:combination}) is an approximation of a linear combination of sparse and dense relevance scores. 
If $p \notin \mathcal{D}_{sp} (\text{or } \mathcal{D}_{ds})$, we directly use the minimum score of $\phi_{sp}(q, p\in \mathcal{D}_{sp})$ \text{or }$\phi_{ds}(q, p\in \mathcal{D}_{ds})$ as a substitute. 
For the sparse and dense retrieval combination, we select the best hyperparameters $\alpha$ (0.1) and $\gamma$ (12) optimizing nDCG@3 on the CAsT19 training set.

\begin{figure}[h!]
\centering
\begin{subfigure}[t]{\columnwidth}
    \includegraphics[clip,width=\columnwidth]{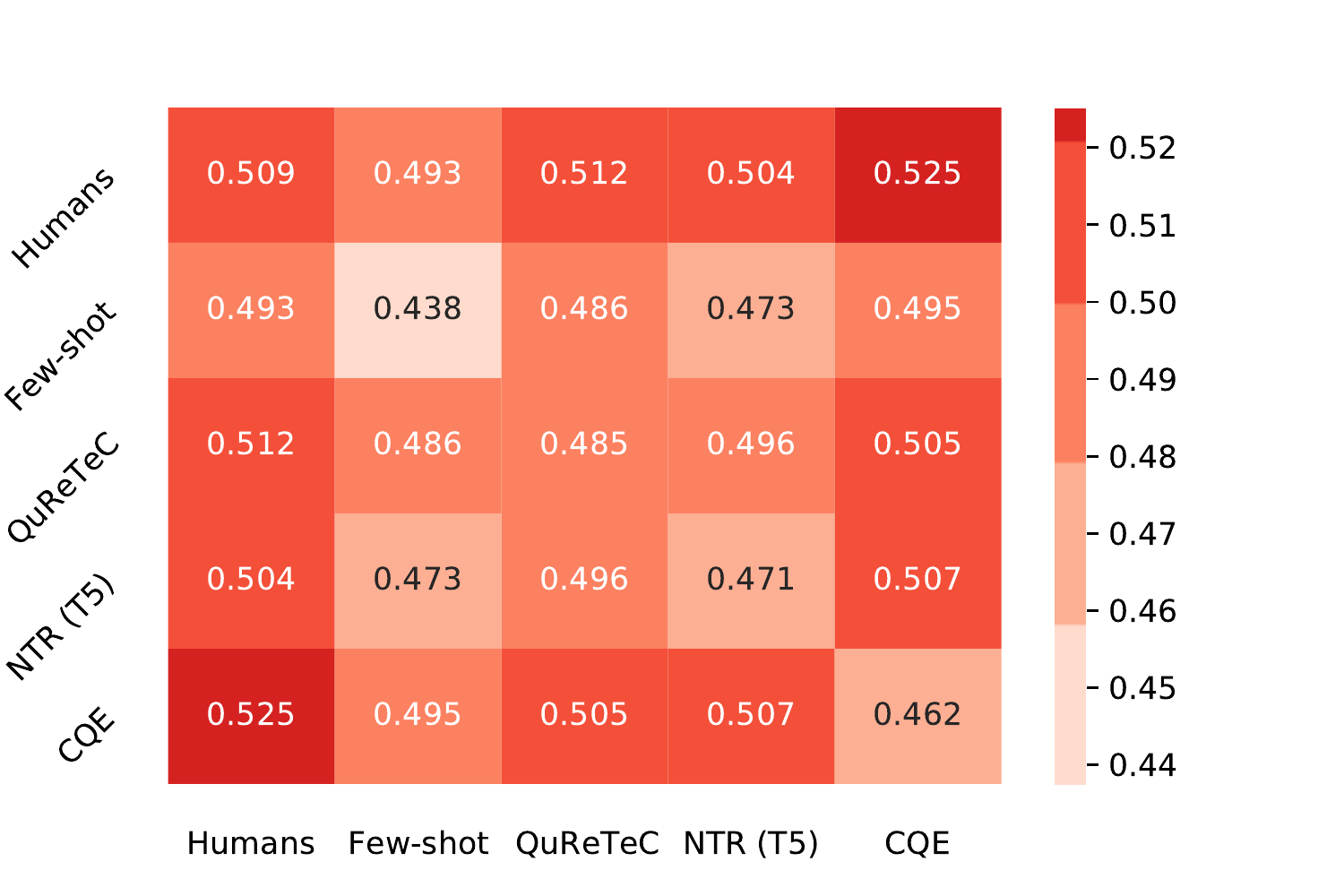}
    \caption{Sparse retrieval fusion}
\end{subfigure}
\begin{subfigure}[t]{\columnwidth}
    \includegraphics[clip,width=\columnwidth]{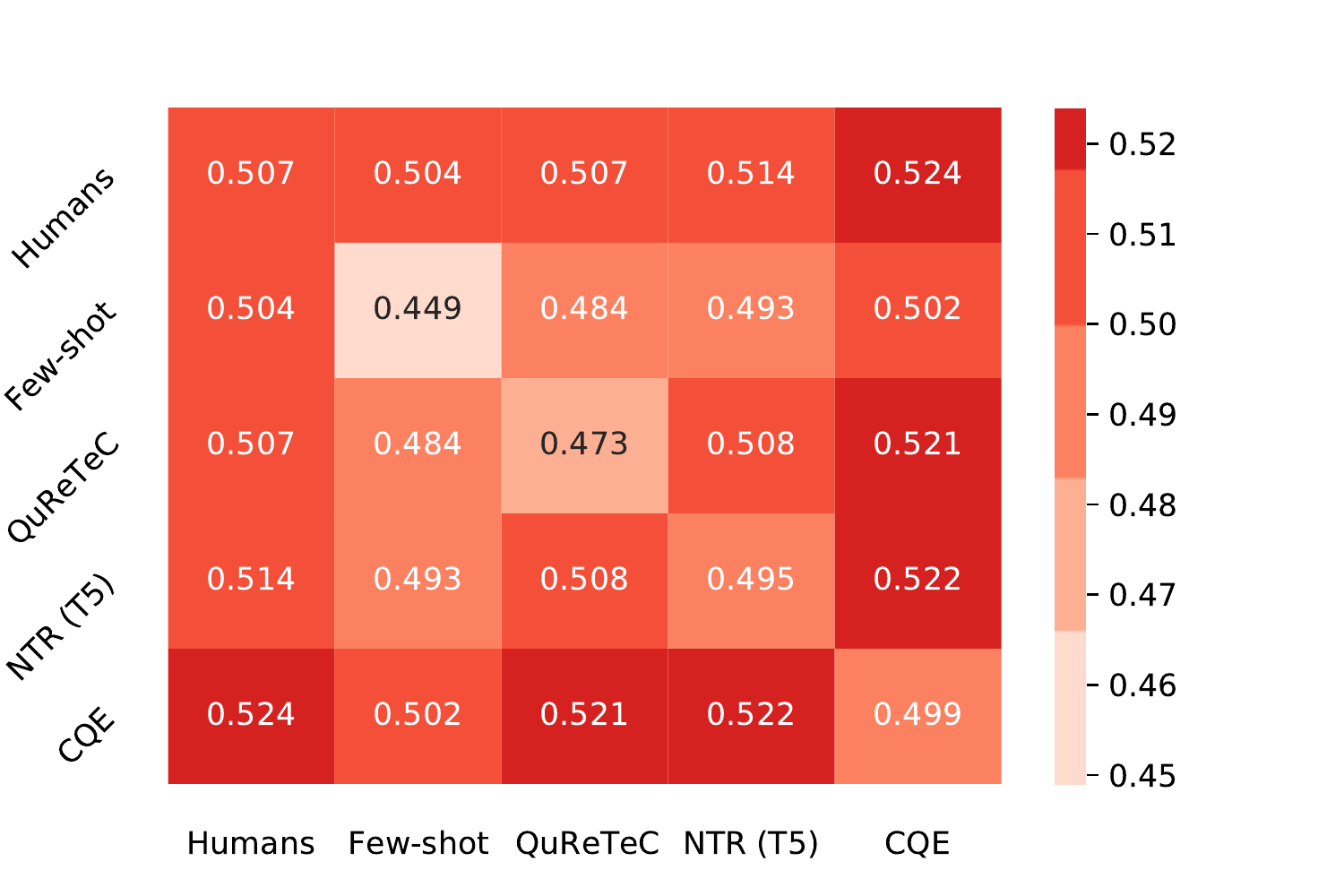}
    \caption{Dense retrieval fusion}
\end{subfigure}
\caption{Ranked list fusion on the CAsT19 eval set, reporting nDCG@3.}
\label{fig:fusion}
\end{figure}

\section{Model Fusion Study}

\noindent We conduct experiments on model fusion to see whether CQE can complement other CQR models in terms of retrieval effectiveness.
Specifically, we use reciprocal rank fusion (RRF) of ranked lists from different queries. 
Figure~\ref{fig:fusion} shows the effectiveness (nDCG@3) of different fusion combinations on the CAsT19 eval set.
We observe that CQE better fuses with all the other CQR models, even in sparse retrieval, where CQE does not perform as well. 
In addition, CQE shows even better fusion results than human queries in dense retrieval.

%% file: project.bbl
\begin{thebibliography}{34}
\expandafter\ifx\csname natexlab\endcsname\relax\def\natexlab#1{#1}\fi

\bibitem[{Ahmad et~al.(2018)Ahmad, Chang, and Wang}]{uddin2018multitask}
Wasi~Uddin Ahmad, Kai-Wei Chang, and Hongning Wang. 2018.
\newblock Multi-task learning for document ranking and query suggestion.
\newblock In \emph{Proc. ICLR}.

\bibitem[{Aliannejadi et~al.(2019)Aliannejadi, Zamani, Crestani, and
  Croft}]{Aliannejadi2019AskingCQ}
Mohammad Aliannejadi, Hamed Zamani, Fabio Crestani, and W.~Bruce Croft. 2019.
\newblock Asking clarifying questions in open-domain information-seeking
  conversations.
\newblock In \emph{Proc. SIGIR}, pages 475--484.

\bibitem[{Choi et~al.(2018)Choi, He, Iyyer, Yatskar, Yih, Choi, Liang, and
  Zettlemoyer}]{quac}
Eunsol Choi, He~He, Mohit Iyyer, Mark Yatskar, Wen-tau Yih, Yejin Choi, Percy
  Liang, and Luke Zettlemoyer. 2018.
\newblock {Q}u{AC}: Question answering in context.
\newblock In \emph{Proc. EMNLP}, pages 2174--2184.

\bibitem[{Dai and Callan(2020)}]{deepct}
Zhuyun Dai and Jamie Callan. 2020.
\newblock Context-aware term weighting for first stage passage retrieval.
\newblock In \emph{Proc. SIGIR}, page 1533–1536.

\bibitem[{Dalton et~al.(2019)Dalton, Xiong, and Callan}]{cast}
Jeffrey Dalton, Chenyan Xiong, and Jamie Callan. 2019.
\newblock {CAsT 2019}: The conversational assistance track overview.
\newblock In \emph{Proc. TREC}.

\bibitem[{Dalton et~al.(2020)Dalton, Xiong, and Callan}]{cast2020}
Jeffrey Dalton, Chenyan Xiong, and Jamie Callan. 2020.
\newblock {CAsT 2020}: The conversational assistance track overview.
\newblock In \emph{Proc. TREC}.

\bibitem[{Elgohary et~al.(2019)Elgohary, Peskov, and Boyd-Graber}]{canard}
Ahmed Elgohary, Denis Peskov, and Jordan Boyd-Graber. 2019.
\newblock Can you unpack that? {L}earning to rewrite questions-in-context.
\newblock In \emph{Proc. EMNLP}, pages 5917--5923.

\bibitem[{Gao et~al.(2020)Gao, Dai, Fan, and Callan}]{clear}
Luyu Gao, Zhuyun Dai, Zhen Fan, and Jamie Callan. 2020.
\newblock Complementing lexical retrieval with semantic residual embedding.
\newblock \emph{arXiv:2004.13969}.

\bibitem[{Hashemi et~al.(2020)Hashemi, Zamani, and Croft}]{guided_transformer}
Helia Hashemi, Hamed Zamani, and W.~Bruce Croft. 2020.
\newblock Guided transformer: Leveraging multiple external sources for
  representation learning in conversational search.
\newblock In \emph{Proc. SIGIR}, page 1131–1140.

\bibitem[{Johnson et~al.(2017)Johnson, Douze, and J{\'e}gou}]{JDH17}
Jeff Johnson, Matthijs Douze, and Herv{\'e} J{\'e}gou. 2017.
\newblock Billion-scale similarity search with {GPUs}.
\newblock \emph{arXiv:1702.08734}.

\bibitem[{Karpukhin et~al.(2020)Karpukhin, Oguz, Min, Lewis, Wu, Edunov, Chen,
  and Yih}]{dpr}
Vladimir Karpukhin, Barlas Oguz, Sewon Min, Patrick Lewis, Ledell Wu, Sergey
  Edunov, Danqi Chen, and Wen-tau Yih. 2020.
\newblock Dense passage retrieval for open-domain question answering.
\newblock In \emph{Proc. EMNLP}, pages 6769--6781.

\bibitem[{Khattab and Zaharia(2020)}]{colbert}
Omar Khattab and Matei Zaharia. 2020.
\newblock {ColBERT}: Efficient and effective passage search via contextualized
  late interaction over {BERT}.
\newblock In \emph{Proc. SIGIR}, page 39–48.

\bibitem[{Kumar and Callan(2020)}]{kumar-callan-2020-making}
Vaibhav Kumar and Jamie Callan. 2020.
\newblock Making information seeking easier: An improved pipeline for
  conversational search.
\newblock In \emph{Proc. EMNLP Findings}.

\bibitem[{Lin et~al.(2021{\natexlab{a}})Lin, Ma, Lin, Yang, Pradeep, and
  Nogueira}]{Lin_etal_SIGIR2021_Pyserini}
Jimmy Lin, Xueguang Ma, Sheng-Chieh Lin, Jheng-Hong Yang, Ronak Pradeep, and
  Rodrigo Nogueira. 2021{\natexlab{a}}.
\newblock {Pyserini}: A {Python} toolkit for reproducible information retrieval
  research with sparse and dense representations.
\newblock In \emph{Proc. SIGIR}, pages 2356--2362.

\bibitem[{Lin et~al.(2020)Lin, Yang, and Lin}]{lin2020distilling}
Sheng-Chieh Lin, Jheng-Hong Yang, and Jimmy Lin. 2020.
\newblock Distilling dense representations for ranking using tightly-coupled
  teachers.
\newblock \emph{arXiv:2010.11386}.

\bibitem[{Lin et~al.(2021{\natexlab{b}})Lin, Yang, and
  Lin}]{lin-etal-2021-batch}
Sheng-Chieh Lin, Jheng-Hong Yang, and Jimmy Lin. 2021{\natexlab{b}}.
\newblock In-batch negatives for knowledge distillation with tightly-coupled
  teachers for dense retrieval.
\newblock In \emph{Proc. RepL4NLP}, pages 163--173.

\bibitem[{Lin et~al.(2021{\natexlab{c}})Lin, Yang, Nogueira, Tsai, Wang, and
  Lin}]{lin2020query}
Sheng-Chieh Lin, Jheng-Hong Yang, Rodrigo Nogueira, Ming-Feng Tsai, Chuan-Ju
  Wang, and Jimmy Lin. 2021{\natexlab{c}}.
\newblock Multi-stage conversational passage retrieval: An approach to fusing
  term importance estimation and neural query rewriting.
\newblock \emph{ACM Trans. Inf. Syst.}, 39(4).

\bibitem[{Nogueira and Cho(2019)}]{marco_BERT}
Rodrigo Nogueira and Kyunghyun Cho. 2019.
\newblock Passage re-ranking with {BERT}.
\newblock \emph{arXiv:1901.04085}.

\bibitem[{Qu et~al.(2020)Qu, Yang, Chen, Qiu, Croft, and Iyyer}]{orconvqa}
Chen Qu, Liu Yang, Cen Chen, Minghui Qiu, W.~Bruce Croft, and Mohit Iyyer.
  2020.
\newblock Open-retrieval conversational question answering.
\newblock In \emph{Proc. SIGIR}.

\bibitem[{Qu et~al.(2019{\natexlab{a}})Qu, Yang, Qiu, Croft, Zhang, and
  Iyyer}]{hae}
Chen Qu, Liu Yang, Minghui Qiu, W.~Bruce Croft, Yongfeng Zhang, and Mohit
  Iyyer. 2019{\natexlab{a}}.
\newblock {BERT} with history answer embedding for conversational question
  answering.
\newblock In \emph{Proc. SIGIR}, page 1133–1136.

\bibitem[{Qu et~al.(2019{\natexlab{b}})Qu, Yang, Qiu, Zhang, Chen, Croft, and
  Iyyer}]{ahs}
Chen Qu, Liu Yang, Minghui Qiu, Yongfeng Zhang, Cen Chen, W.~Bruce Croft, and
  Mohit Iyyer. 2019{\natexlab{b}}.
\newblock Attentive history selection for conversational question answering.
\newblock In \emph{Proc. CIKM}, page 1391–1400.

\bibitem[{Quan et~al.(2019)Quan, Xiong, Webber, and Hu}]{quan-etal-2019-gecor}
Jun Quan, Deyi Xiong, Bonnie Webber, and Changjian Hu. 2019.
\newblock {GECOR}: An end-to-end generative ellipsis and co-reference
  resolution model for task-oriented dialogue.
\newblock In \emph{Proc. EMNLP}, pages 4547--4557.

\bibitem[{Radlinski and Craswell(2017)}]{conv_search}
Filip Radlinski and Nick Craswell. 2017.
\newblock A theoretical framework for conversational search.
\newblock In \emph{Proc. CHIIR}, pages 117--126.

\bibitem[{Rastogi et~al.(2019)Rastogi, Gupta, Chen, and Lambert}]{scaling}
Pushpendre Rastogi, Arpit Gupta, Tongfei Chen, and Mathias Lambert. 2019.
\newblock Scaling multi-domain dialogue state tracking via query reformulation.
\newblock In \emph{Proc. NAACL}, pages 97--105.

\bibitem[{Reddy et~al.(2019)Reddy, Chen, and Manning}]{coqa}
Siva Reddy, Danqi Chen, and Christopher~D. Manning. 2019.
\newblock {CoQA}: A conversational question answering challenge.
\newblock \emph{Transactions of the Association for Computational Linguistics},
  pages 249--266.

\bibitem[{Reimers and Gurevych(2019)}]{sentence-bert}
Nils Reimers and Iryna Gurevych. 2019.
\newblock {Sentence-BERT}: Sentence embeddings using siamese {BERT}-networks.
\newblock In \emph{Proc. ACL}.

\bibitem[{Ren et~al.(2018)Ren, Ni, Malik, and Ke}]{cqu}
Gary Ren, Xiaochuan Ni, Manish Malik, and Qifa Ke. 2018.
\newblock Conversational query understanding using sequence to sequence
  modeling.
\newblock In \emph{Proc. WWW}, pages 1715--1724.

\bibitem[{Vakulenko et~al.(2020)Vakulenko, Longpre, Tu, and
  Anantha}]{vakulenko2020question}
Svitlana Vakulenko, Shayne Longpre, Zhucheng Tu, and Raviteja Anantha. 2020.
\newblock Question rewriting for conversational question answering.
\newblock \emph{arXiv:2004.14652}.

\bibitem[{Voskarides et~al.(2019)Voskarides, Li, Panteli, and Ren}]{ilips}
Nikos Voskarides, Dan Li, Andreas Panteli, and Pengjie Ren. 2019.
\newblock {ILPS} at {TREC} 2019 conversational assistant track.
\newblock In \emph{Proc. TREC}.

\bibitem[{Voskarides et~al.(2020)Voskarides, Li, Ren, Kanoulas, and
  de~Rijke}]{quretec}
Nikos Voskarides, Dan Li, Pengjie Ren, Evangelos Kanoulas, and Maarten
  de~Rijke. 2020.
\newblock Query resolution for conversational search with limited supervision.
\newblock In \emph{Proc. SIGIR}, pages 921--930.

\bibitem[{Xiong et~al.(2021)Xiong, Xiong, Li, Tang, Liu, Bennett, Ahmed, and
  Overwijk}]{xiong2020approximate}
Lee Xiong, Chenyan Xiong, Ye~Li, Kwok-Fung Tang, Jialin Liu, Paul~N. Bennett,
  Junaid Ahmed, and Arnold Overwijk. 2021.
\newblock Approximate nearest neighbor negative contrastive learning for dense
  text retrieval.
\newblock In \emph{Proc. ICLR}.

\bibitem[{Yang et~al.(2019{\natexlab{a}})Yang, Lin, Wang, Lin, and Tsai}]{hqae}
Jheng{-}Hong Yang, Sheng{-}Chieh Lin, Chuan{-}Ju Wang, Jimmy Lin, and
  Ming{-}Feng Tsai. 2019{\natexlab{a}}.
\newblock Query and answer expansion from conversation history.
\newblock In \emph{Proc. TREC}.

\bibitem[{Yang et~al.(2019{\natexlab{b}})Yang, Xie, Lin, Li, Tan, Xiong, Li,
  and Lin}]{bertserini}
Wei Yang, Yuqing Xie, Aileen Lin, Xingyu Li, Luchen Tan, Kun Xiong, Ming Li,
  and Jimmy Lin. 2019{\natexlab{b}}.
\newblock End-to-end open-domain question answering with {BERT}serini.
\newblock In \emph{Proc. ACL (Demonstrations)}, pages 72--77.

\bibitem[{Yu et~al.(2020)Yu, Liu, Yang, Xiong, Bennett, Gao, and
  Liu}]{yu2020fewshot}
Shi Yu, Jiahua Liu, Jingqin Yang, Chenyan Xiong, Paul Bennett, Jianfeng Gao,
  and Zhiyuan Liu. 2020.
\newblock Few-shot generative conversational query rewriting.
\newblock In \emph{Proc. SIGIR}, pages 1933--1936.

\end{thebibliography}
